# The Impact of Socioeconomic Factors on Health Disparities

**Krish Khanna***

University of Chicago Laboratory Schools

**Jeffrey Lu***

York Community High School

**Jay Warrier***

Khan Lab School

* Equal contribution and shared first authorship

**Note that the quality of this paper is limited by the lack of a complete high school education of all authors.**

*High-quality healthcare in the US can be cost-prohibitive for certain socioeconomic groups. In this paper, we examined data from the US Census and the CDC to determine the degree to which specific socioeconomic factors correlate with both specific and general health metrics. We employed visual analysis to find broad trends and predictive modeling to identify more complex relationships between variables. Our results indicate that certain socioeconomic factors, like income and educational attainment, are highly correlated with aggregate measures of health.*

## 1.    Introduction

Currently, the United States healthcare system has a "cruel tendency to delay or deny high-quality care to those who are most in need of it but can least afford its high cost," (Shmerling) resulting in rampant disparities in health outcomes throughout the nation. The news of today is riddled with stories of people receiving poor care due to systematic biases present in the modern healthcare system and the effect of the increasingly unaffordable cost of life-saving medication. In order to better understand the degree to which this inequality exists, we investigated which socioeconomic indicators model health outcomes best.

## 2.    Related Works

In his 2015 report for the Urban Institute titled, "How are Income and Wealth Tied to Health and Longevity," Dr. Stephen H. Woolf showed a negative correlation between income and health issues, analyzing both subjective survey data along with disease prevalence across income groups. Additionally, they formulated different justifications on why this could be the case, mentioning that richer people could afford a healthier lifestyle, while poorer people usually work more dangerous jobs with less access to health insurance and therefore lower-quality care.

In their 2004 article for Academic Medicine, Fiscella and Williams analyze "Health Disparities Based on Socioeconomic Inequities" in the context of urban medical practice. From the outset, they define socioeconomic status as something that, "whether measured by income, educational achievement, or occupation" and is "associated with large disparities in health status." The article argues that there is a strong causal relationship between socioeconomic factors and health outcomes, and supports it with qualitative as well as data-driven arguments. Additionally, the authors claim that low income and bad health outcomes combine to form a self-perpetuating cycle.

## 3. Initial Questions

Initially, we sought to discover if and how socioeconomic factors correlate with the quality of healthcare in the US, but given the qualitative and subjective nature of healthcare quality, we decided to refine our core question. Ultimately, we decided to examine health *outcomes* because it's directly influenced by the quality of healthcare. We also wanted to investigate how we could address any disparities we do find. We hypothesized that race, income, region, and education would have the most influence on health outcomes and that the major determinants of successful health outcomes would be mostly immutable socioeconomic factors.

## 4. Dataset Description and Cleaning

We used 3 datasets containing socioeconomic, educational attainment, and healthcare data from the US Census and the CDC. It should be noted that this data is from 2017, which means COVID is not a factor in any of our analyses, and models may not generalize well to the post-COVID US. Each of our datasets was organized by tract, which is a sub-county division of land used by the US government. The socioeconomic dataset included tract distributions of income, job sector, commute type, race, and gender. The education dataset specified the proportion of tract populations who had attained a specific level of education, such as a high school diploma or college degree. It also included limited economic status data and measures for each level of education. (i.e. percentage of people who attained a high school diploma that are now in poverty) Finally, the healthcare data included the prevalence of specific health conditions, such as arthritis and depression, and aggregated measures of general physical and mental health.

Because most of our data was sourced directly from the US government without any processing, we needed to clean and prepare the data before we could use it. The datasets had various issues that prevented us from using the raw data, including excessively long column names, inconsistent null values, and missing data. Our approach involved first cleaning each dataset individually, then combining them for final cleaning. One especially challenging dataset was the education dataset. Because there were over 70,000 rows to begin with, we had to identify columns with useful data and import only those columns. Then, we put it through the same procedure as all the other datasets. This involved removing rows with null values and standardizing column names. Once this was done, we combined all of them for final cleaning. In all, we only discarded about 2,000 tracts of data. Although it would be ideal to use all tracts, this is an acceptable trade-off because we still have a large and diverse dataset that represents an entire continuum of communities with different socioeconomic factors.

## 5. Exploratory Analysis

Initially, we looked at a heatmap of all of the variables (Figure 5.1) to find linear correlations. However, this proved to be quite difficult to interpret given the immense amount of variables we had. Because of this, we decided to test our hypothesis by exploring the direct relationships between variables that we intuitively believed could be correlated and showed a promising correlation coefficient.

We started by visualizing the relationship between income and bad physical health (Figure 5.2). In accordance with our hypothesis and related works, there was a negative correlation between income and physical health. Interestingly, this graph showed a visually strong nonlinear correlation, with the rate of change getting increasingly slower further across the x-axis. There appears to be a significant gap between low and middle-income tracts, but little difference between middle and high-income tracts. We will explore this trend further in the predictive analysis.

Next, we decided to explore education's effect on health. We found that high school graduation is an important milestone for an individual.(Figure 5.3) Because of this, we decided to model the difference in physical health between tracts with a high percentage of high school graduates (>70%) and tracts with a low percentage of graduates (<70%). We see a noticeable difference between the two groups, with the high percentage group having an average of 10% less of their population in bad physical health.

We also explored race and region's effect on health. When investigating regional differences, (Figure 5.4) the Midwest and South seemed to have slightly worse physical health than the Northeast and West, but these were very small differences. In our race graphs, (Figure 5.5) we saw slight trends, as the percentage of Caucasian people was inversely correlated with bad physical health, and the opposite was true for most minorities. However, we noticed an irregular relationship between Asian American proportions and bad physical health, with the linear model fitting very poorly to the curved distribution. While the heteroskedasticity of the data leaves us unable to make a valid conclusion from the linear model, this relationship is something we could explore in the future.

Lastly, we explored the impact of specific health indicators on outcomes. We found that some of the indicators were very promising and had high correlation coefficients with health indicators (Figure 6.6). These indicators were regular checkups, screening rates, smoking, sleep, and physical activity.

## 6. Predictive Analysis

Our objective was to determine the impact of socioeconomic factors on health outcomes in general. We decided that, in order to best utilize our wealth of data, we would analyze both specific disorder outcomes, such as depression and arthritis, as well as aggregate level indicators, which were the total prevalence of bad physical and mental health in a tract.

First, we tried to use simple linear regression to model basic relationships between individual variables. The model's objective was to fit a line through data points to minimize the mean squared error, which we believed would be useful to determine potential variables for further investigation as well as test the robustness of improved future models. Although the relationships between some variables were clearly linear, others were heteroskedastic or had nonlinear trends (Figures 6.1).

In order to better explore the nonlinear trends, we decided to employ single-variable polynomial regression with degrees up to 4. This model seemed to perform very well both visually and quantitatively, but the data is actually extremely heteroskedastic (Figure 6.2). We also weren't able to identify with certainty all of the most revealing indicators because we could only examine one feature at a time.

To address this issue, we decided to build regression tree models (Figure 6.3) for each of the different health outcomes. The regression tree model has three major fundamental advantages over the previous models we had built: explainability, higher input dimensionality, and input feature prioritization. Because this model optimizes which inputs it chooses to look at, we fed all the available data into our model. To maintain explainability, we set the maximum depth very low.

Our model used a binary tree structure with variance for determining splits. Every training point starts in the root node and is split into child nodes at every non-leaf node. Our splits were chosen across a single axis $j$ and a constant threshold $t$ for every node $v$ with child nodes $v_0$ and $v_1$, with $S_v$ representing all points in node $v$:

$$S_{v_0}^{(j,t)} = \{(x^{(i)}, y^{(i)}) \in S_v \mid x_j^{(i)} < t\} \qquad S_{v_1}^{(j,t)} = \{(x^{(i)}, y^{(i)}) \in S_v \mid x_j^{(i)} \geq t\}$$

Variance for each of the child nodes is defined as mean squared error from the mean:

$$V_{v_0}^{(j,t)} = \frac{1}{|S_{v_0}^{(j,t)}|} \sum_{(x^{(i)}, y^{(i)}) \in S_{v_0}^{(j,t)}} (y^{(i)} - \frac{\sum_{(x^{(k)}, y^{(k)}) \in S_{v_0}^{(j,t)}} (y^{(k)})}{|S_{v_0}^{(j,t)}|})^2$$

$$V_{v_1}^{(j,t)} = \frac{1}{|S_{v_1}^{(j,t)}|} \sum_{(x^{(i)}, y^{(i)}) \in S_{v_1}^{(j,t)}} (y^{(i)} - \frac{\sum_{(x^{(k)}, y^{(k)}) \in S_{v_1}^{(j,t)}} (y^{(k)})}{|S_{v_1}^{(j,t)}|})^2$$

The loss for a split with some axis and threshold is the sum of the variances in each of the child nodes:

$$L(S_v, j, t) = (V_{v_0}^{(j,t)}) + (V_{v_1}^{(j,t)})$$

At every node, the threshold $t$ and axis $j$ are determined by minimizing the loss. Once all the splits are determined, the model's prediction for all points split into some leaf node $v$ is the value of $y^*$ that minimizes the loss function of the leaf for all the training values, which is the mean of all the y values in the leaf node (Proof 0.1):

$$\hat{y}_v = \arg\min_{y^*} \frac{1}{|S_v|} \sum_{(x^{(i)}, y^{(i)}) \in S_v} (y^{(i)} - y^*)^2 = \frac{1}{|S_v|} \sum_{(x^{(i)}, y^{(i)}) \in S_v} y^{(i)}$$

Although this model is highly explainable and performed surprisingly well with an $R^2$ score of 0.642, it lacks the ability to extrapolate to values that are out of the range of what it was trained on and may still overfit to the data. We showed that every decision tree overfits after the max depth was increased beyond a certain threshold. (Figure 6.4) To reduce these errors, we decided to employ a random forest regression model. This decreased the explainability of the model but had a significant positive impact on the accuracy.

Each forest was composed of a number of trees, with each tree generated on a random subset of the data. The final prediction of this model is the mean of the predictions of all the trees. This increases the performance by reducing errors from any individual tree. We also found that this was an effective method of reducing overfitting. (Figures 6.4 and 6.5)

We ran this model on two sets of inputs: one containing socioeconomic factors and another containing health statistic data, such as percent up-to-date with screenings and percent smoking. All socioeconomic factors and the four most correlated health indicators were used in the model.(Figure 6.6) To our surprise, the health indicators produced models with significantly higher accuracy (~0.92 $R^2$ on aggregate indicators and $0.7 < R^2 < 0.9$ on specific indicators) than the socioeconomic factors (~0.84 $R^2$ on aggregate indicators and $0.6 < R^2 < 0.8$ on specific indicators). This is especially interesting as the health indicators are simultaneously addressable and not directly correlated with income (~0.7 $R^2$) when predicting income with the same model and factors.

It's known that models can achieve very high accuracies by overfitting instead of learning, particularly models like decision trees. In order to ensure that our models aren't overfitting, we implemented cross-validation on our best-performing predictors for all the health variables. Each model was able to extrapolate to unseen data, with $R^2$ scores up to 0.845, even when a significant portion of its training data was hidden. (Figure 6.7) Additionally, the distribution for the scores of the models had very little spread, meaning the model complexity is a good fit for the data.

## 7.   Ethical Considerations

Given that we are utilizing sensitive healthcare data, there are many ethical questions that arise. First and foremost, using healthcare data at all can be seen as a major breach of privacy. Since we used data aggregated by tract, however, we still have valuable information without ever knowing the health of any individual. Another potential ethical issue could be using data that doesn't reflect the true state of America, whether it is because of null values or unreported cases, as it could skew the results away from the true underlying distribution. Although this is a valid concern, it's impossible to obtain perfect data and we're working with data directly from the government, which is the best we can find. Lastly, models trained on race and gender data could make the problematic assertion that race and gender determine health. Fortunately, our best-performing models weren't very influenced by these factors from the start. It's still important, however, to emphasize that all trends found are correlations and that finding causational relationships between race, gender, and health outcomes is beyond the scope of this work.

Khanna, Lu, Warrier - 6

## 8. Conclusion

Overall, education and income both proved to be good predictors of health, but race and region turned out to be the opposite. Our best-performing model on socioeconomic data, the random forest regressor, ended up with good performance (0.845 $R^2$) in aggregate health measures with reduced performance (~0.660-0.752 $R^2$) in specific health measures, a trend reflected across almost all of our model types. In contrast, we found incredibly strong relationships between certain health indicators and outcomes that performed significantly better on both aggregate and specific outcomes without being collinear with socioeconomic factors, with the best model ending with an impressive highest performance (0.92 $R^2$) and good performance for specific outcomes (~0.7-0.9 $R^2$). Our analyses showed that, generally, socioeconomic factors can predict the *overall* health of a tract, but cannot predict the *specific* health issues. Although we wanted to create an explainable model with high performance, we could only use lower-complexity models to guide the creation of other models. This failure proves that the relationship between socioeconomic factors and health outcomes is very complicated and cannot be distilled into a few simple rules.

Our results show that socioeconomic factors, specifically income and level of educational attainment, correlate with certain health outcomes through visual analysis and robust predictive modeling. Additionally, we found that these disparities are even more correlated with certain "fixable problems," which provides a potential pathway for further research to investigate how the disparities present in the healthcare system today can be addressed.

## 9. Acknowledgements

This paper was originally written for STAT 10118 in 2022 at the University of Chicago and was later slightly revised for publication. We thank our instructors Phillip Lo and Adela DePavia as well as our TA Ray Fregly for guidance on all aspects of this work.

## 10. Revisions

The original version of this paper erroneously listed Phillip Lo, Adela DePavia, and Ray Fregly as authors. This revision of the paper has corrected this error.

## Appendix:
## Figure 5.1 - Heatmap of Correlations Between All Variables:

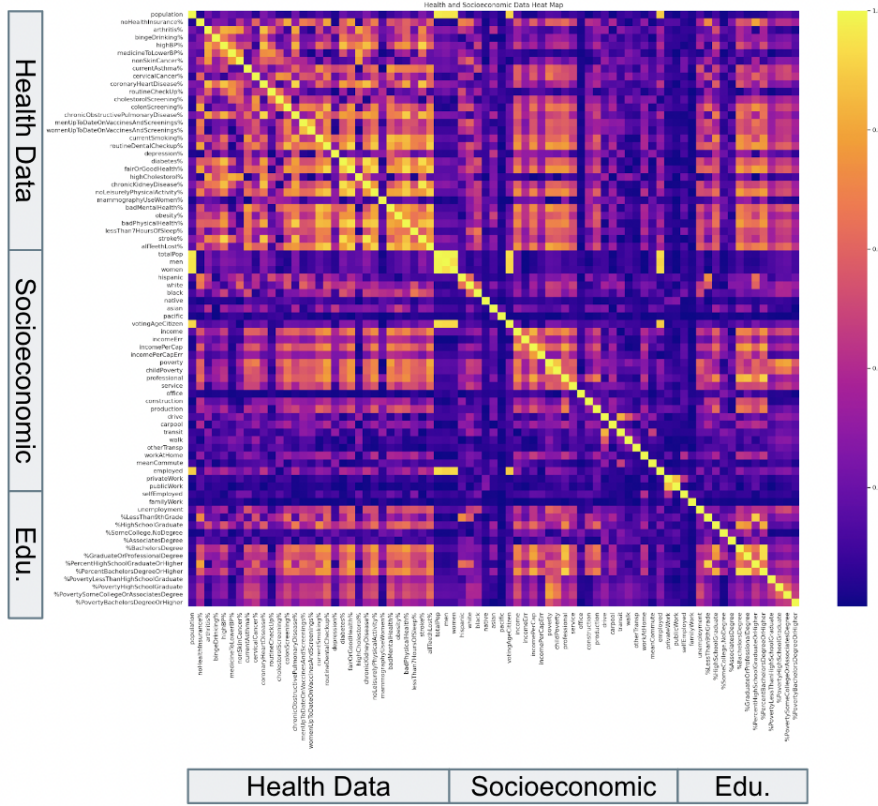

## Figure 5.2 - Income vs. Physical Health Scatterplot:

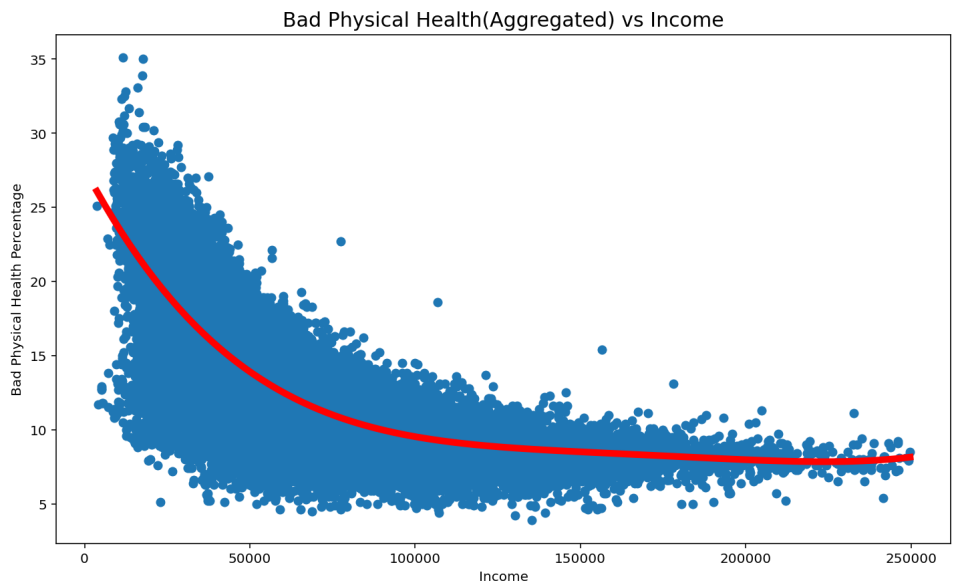

**Figure 5.3 - Physical Health vs. High School Graduation Histogram:**

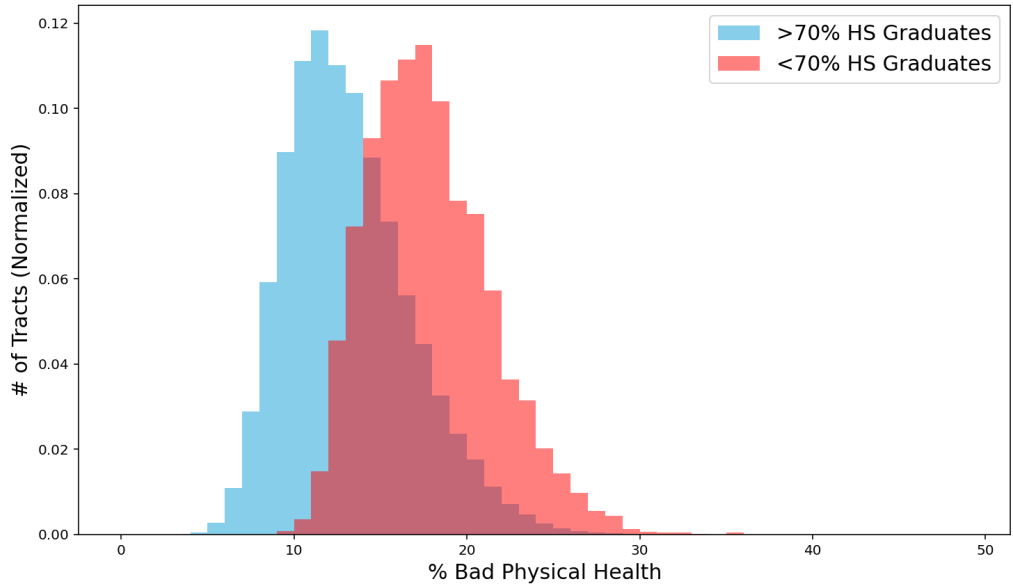

**Figure 5.4 - Physical Health Distribution in Different Regions**

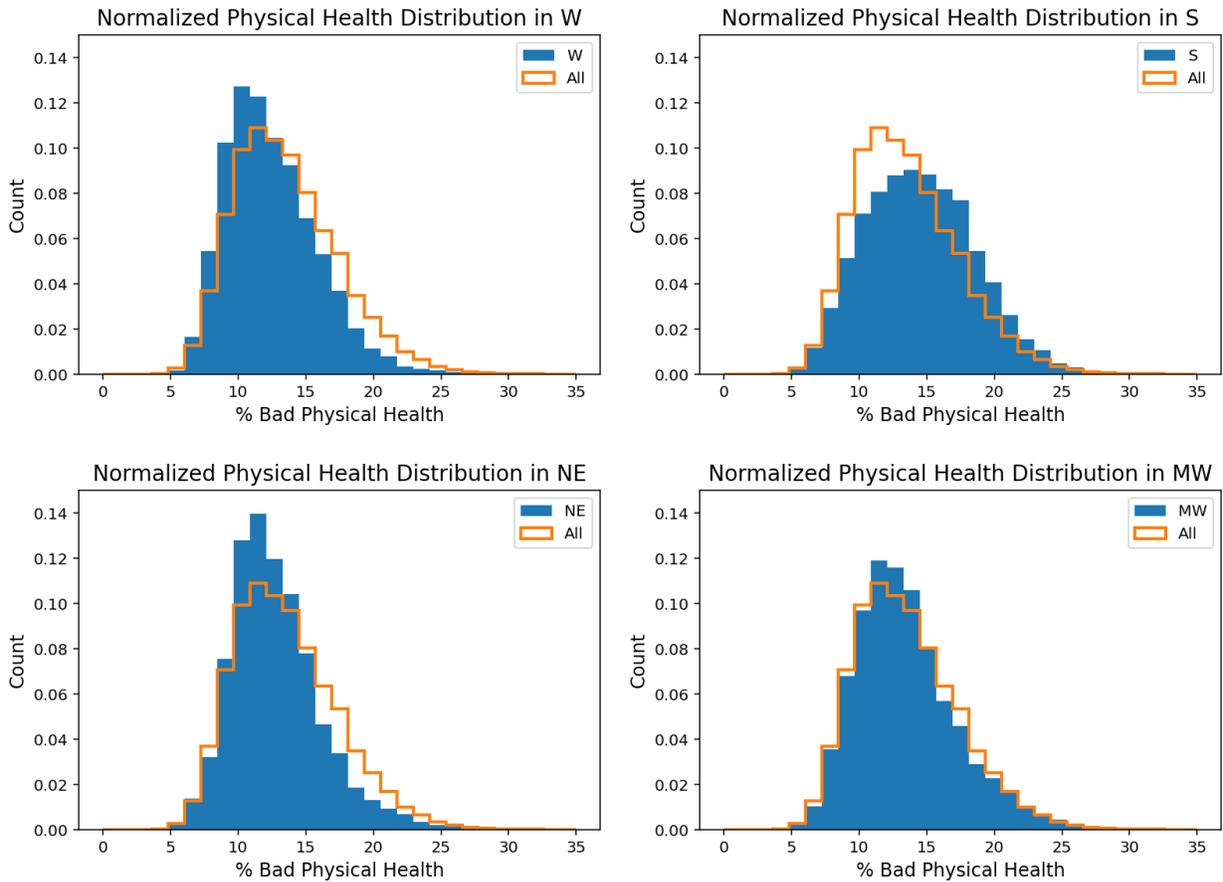

## Figure 5.5 - Race and Physical Health

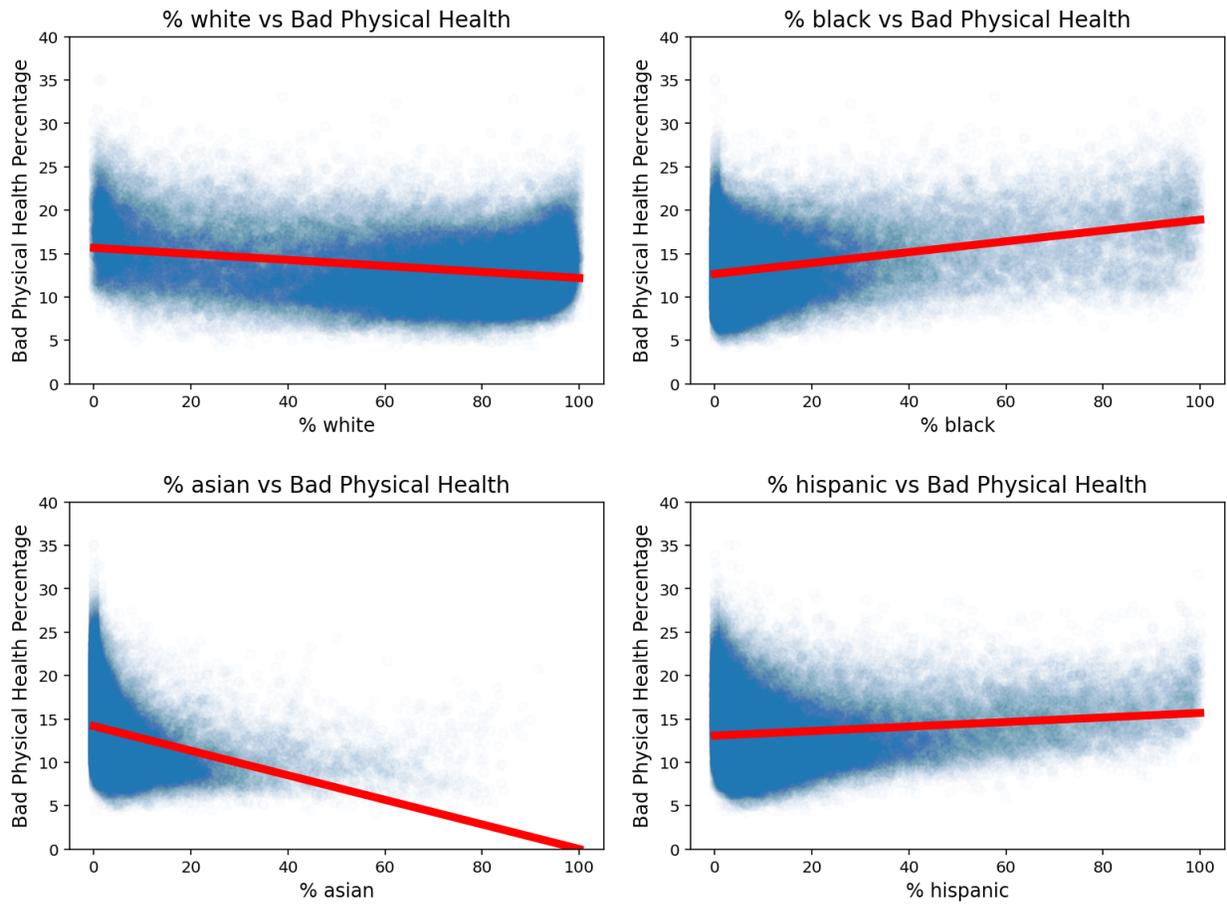

## Figure 6.1 - Linear Modeling:
## Figure 6.1.1 - Linear Model to Predict Cancer Rates based on Poverty:

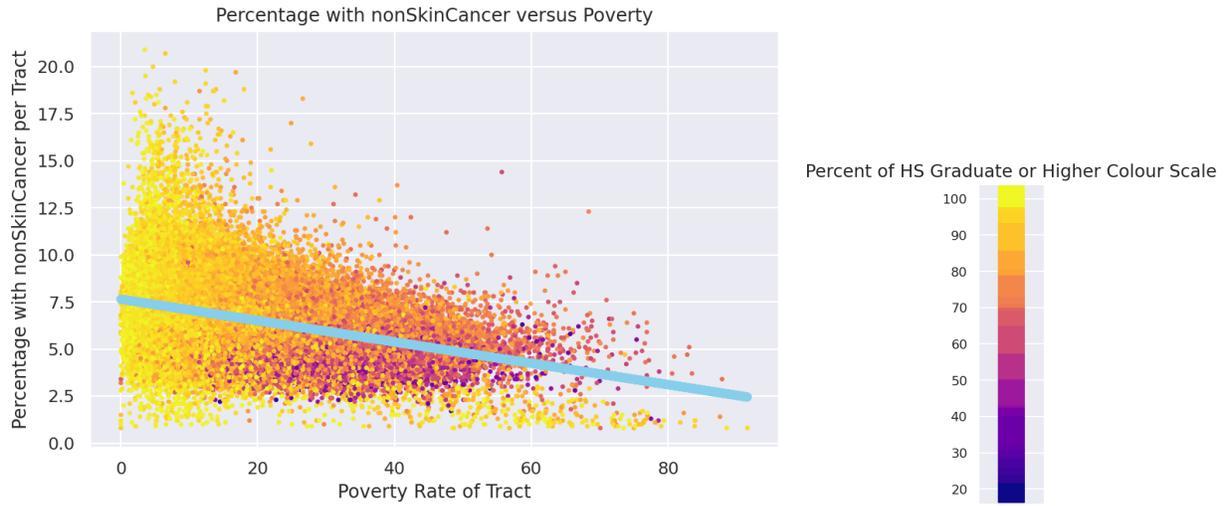

## Figure 6.1.2 - Residuals of the linear model in Figure 6.1:

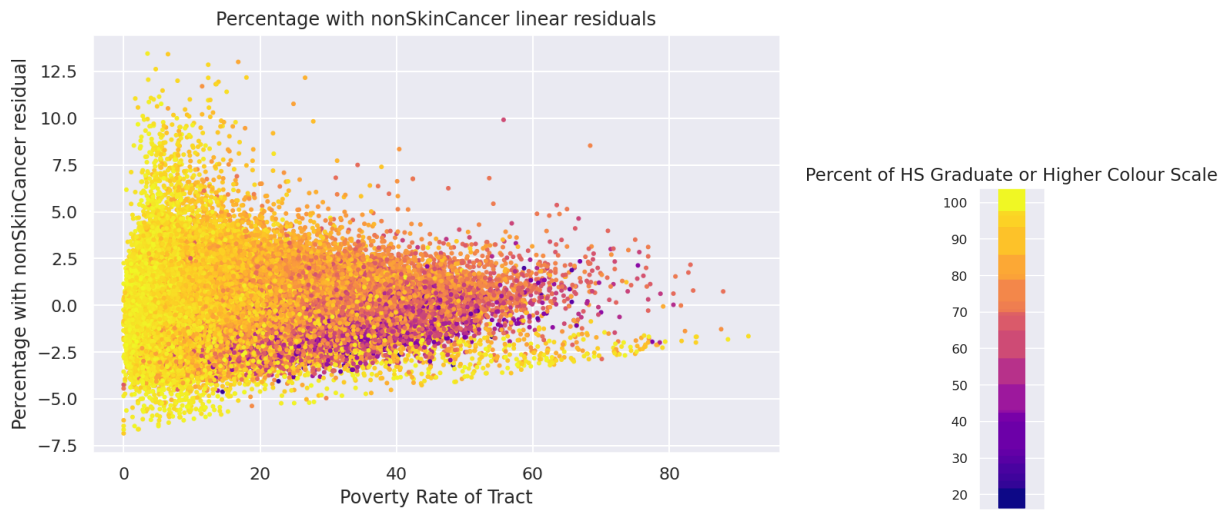

## Figure 6.2 - Heteroskedasticity in visually satisfactory model

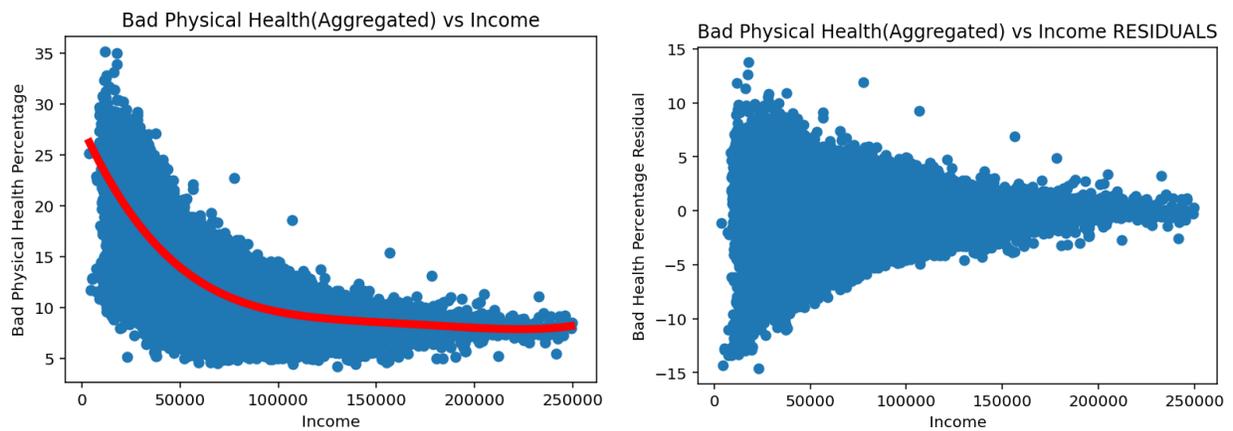

**Figure 6.3 - "Simple" Decision Tree Example (Predicting badMentalHealth%)**

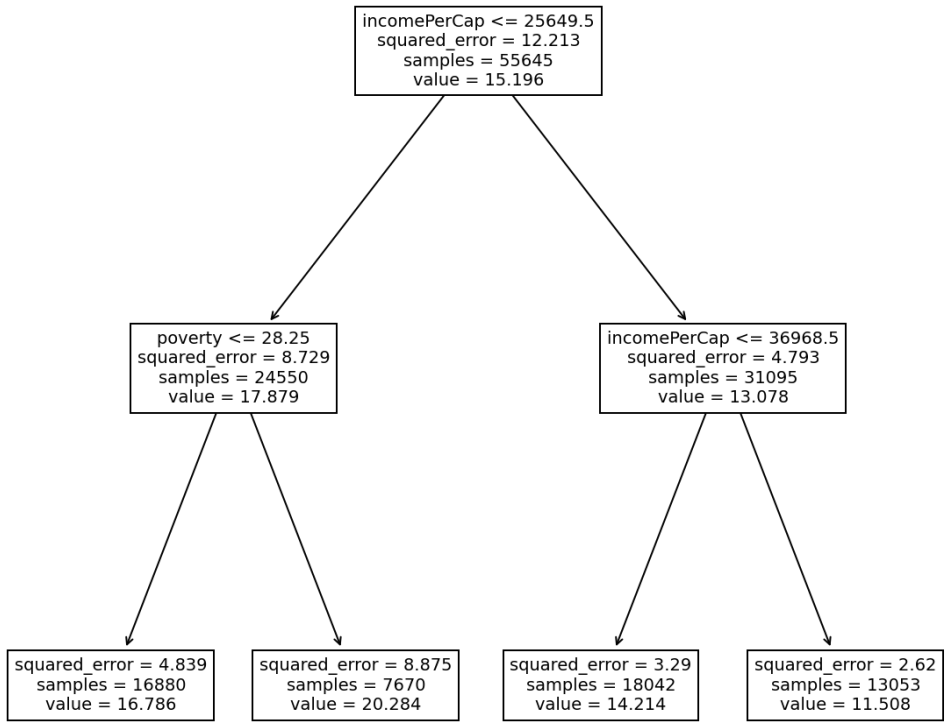

**Figure 6.4 - Finding overfit threshold for every decision tree**
**NOTE:** Red is testing metrics; Blue is training metrics

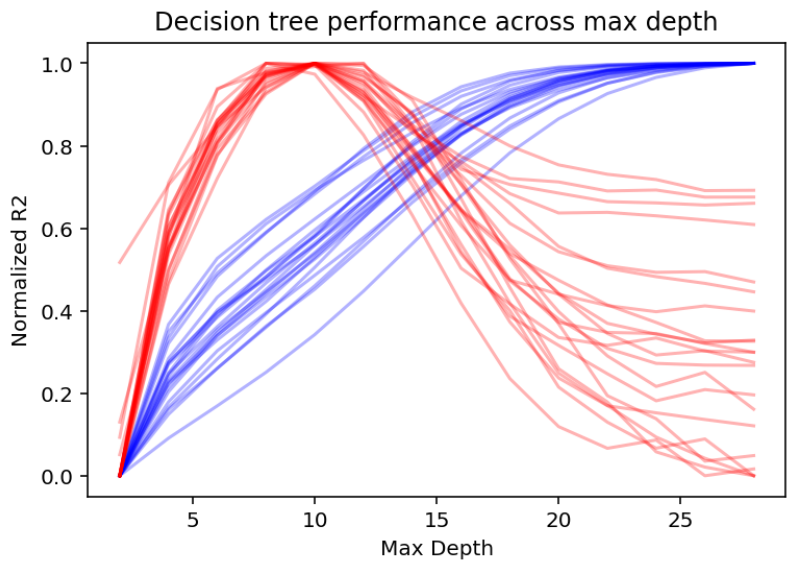

**Figure 6.5 - Repeating graph from Figure 6.4 but with Random Forest**
**NOTE:** Red is testing metrics; Blue is training metrics

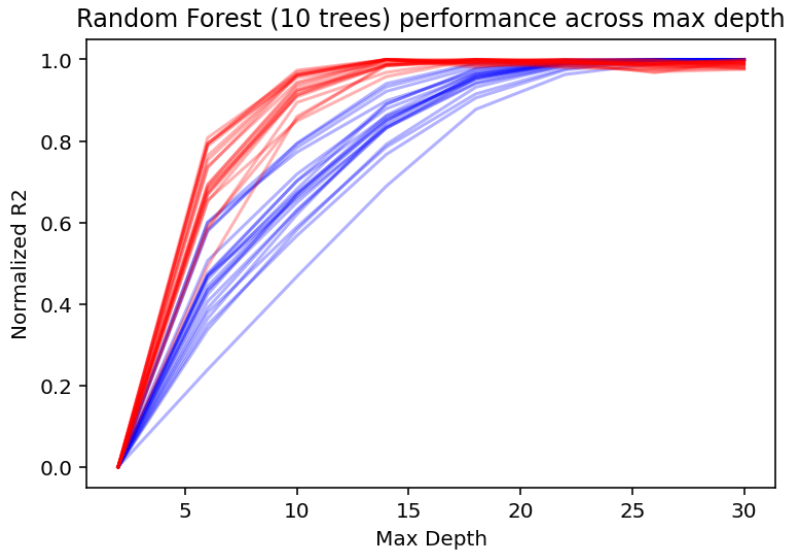

**Figure 6.6 - Correlations between "fixable problems" and health indicators**

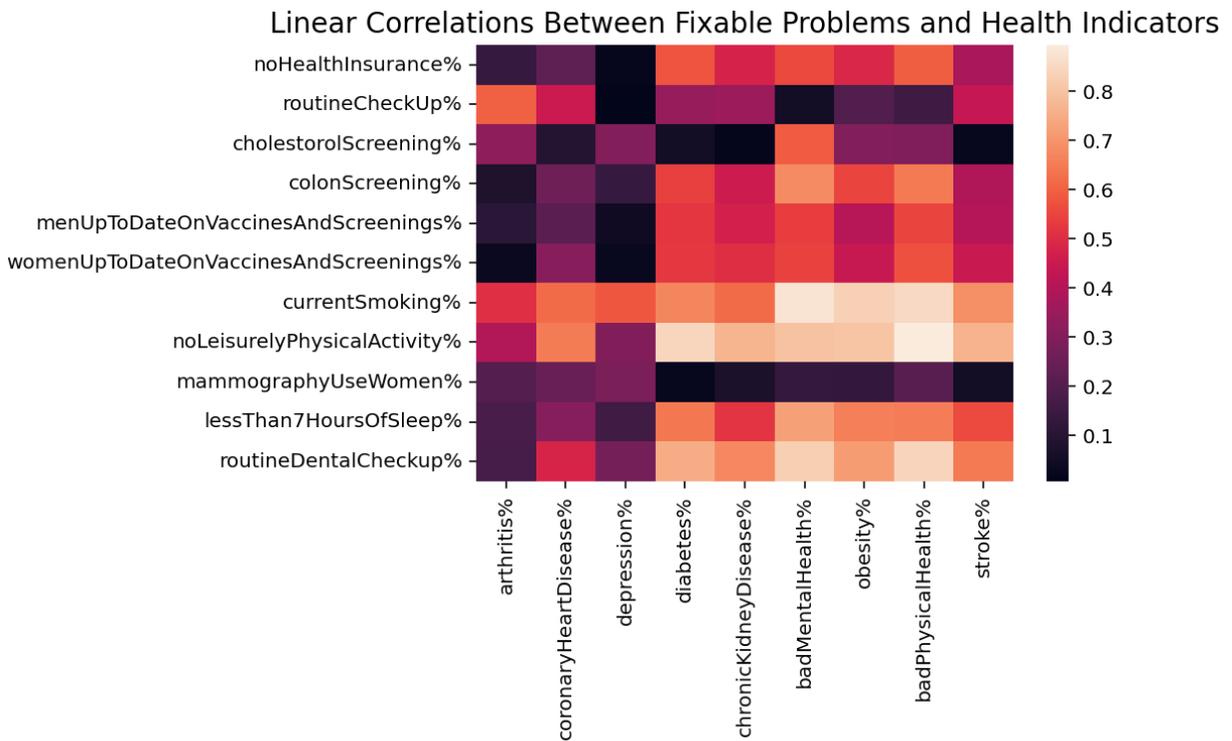

## Figure 6.7 - Examples of score distributions from random forest with 5 fold cross-validation

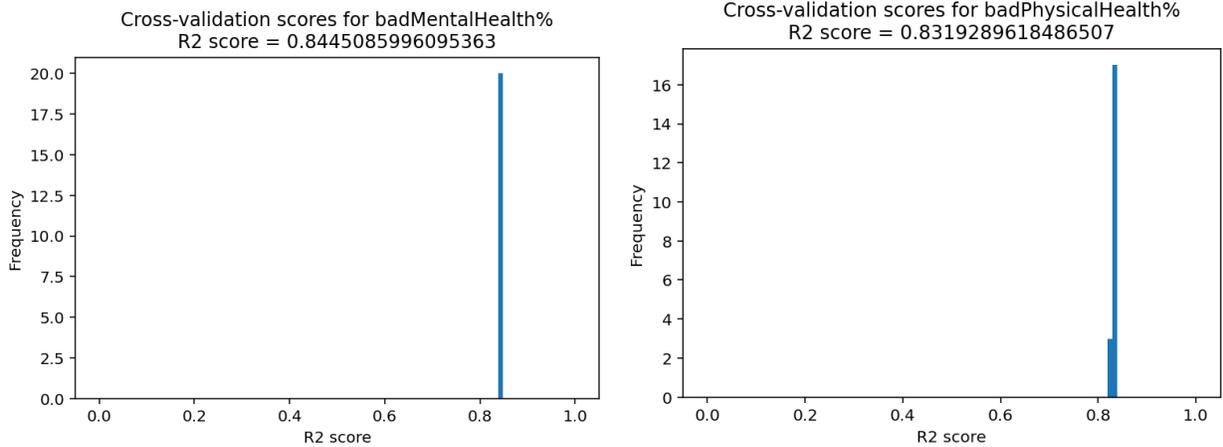

**Proof 0.1:**

We seek to prove:

$$\arg\min_{y^*} \frac{1}{|S_v|} \sum_{(x^{(i)},y^{(i)}) \in S_v} (y^{(i)} - y^*)^2 = \frac{1}{|S_v|} \sum_{(x^{(i)},y^{(i)}) \in S_v} y^{(i)}$$

We can take the derivative of the LHS of the equation to identify critical points:

$$\frac{d}{dy^*}\left(\frac{1}{|S_v|} \sum_{(x^{(i)},y^{(i)}) \in S_v} (y^{(i)} - y^*)^2\right)$$

Using the Constant Multiple Rule and the Sum Rule:

$$= \frac{1}{|S_v|} \sum_{(x^{(i)},y^{(i)}) \in S_v} \frac{d}{dy^*}\left((y^{(i)} - y^*)^2\right)$$

By Chain Rule:

$$= \frac{1}{|S_v|} \sum_{(x^{(i)},y^{(i)}) \in S_v} (2(y^* - y^{(i)}))$$

Factoring out the 2 and setting this equal to 0 to find the critical point:

$$\frac{2}{|S_v|} \sum_{(x^{(i)},y^{(i)}) \in S_v} (y^* - y^{(i)}) = 0$$

Dividing both sides by $\frac{2}{|S_v|}$:

$$\sum_{(x^{(i)},y^{(i)})\in S_v} (y^* - y^{(i)}) = 0$$

Expanding the summation:

$$\sum_{(x^{(i)},y^{(i)})\in S_v} y^* - \sum_{(x^{(i)},y^{(i)})} y^{(i)} = 0$$

Isolating sums on different sides of the equation:

$$\sum_{(x^{(i)},y^{(i)})\in S_v} y^* = \sum_{(x^{(i)},y^{(i)})} y^{(i)}$$

Because $y^*$ is a constant across all points in a leaf node, we can treat it like a constant and replace the sum by multiplying by the number of elements:

$$|S_v| y^* = \sum_{(x^{(i)},y^{(i)})} y^{(i)}$$

Dividing by $|S_v|$, we arrive at the conclusion:

$$y^* = \frac{1}{|S_v|} \sum_{(x^{(i)},y^{(i)})} y^{(i)}$$

To prove that this is a minimum, we must show:

$$\frac{d}{dy^*}\left[\frac{2}{|S_v|} \sum_{(x^{(i)},y^{(i)})\in S_v} (y^* - y^{(i)})\right] > 0$$

Using the Constant Multiple Rule and the Sum Rule on LHS:

$$LHS = \frac{2}{|S_v|} \sum_{(x^{(i)},y^{(i)})\in S_v} \left(\frac{d}{dy^*}(y^*) - \frac{d}{dy^*}(y^{(i)})\right)$$

Using the Constant Rule and the Power Rule:

$$= \frac{2}{|S_v|} \sum_{(x^{(i)},y^{(i)})\in S_v} (1 - 0)$$

Simplifying the summation:

$$= \frac{2}{|S_v|} \times |S_v| = 2$$

Returning to our original inequality with our simplified LHS:

$$2 > 0$$

It is clear that the inequality is true, so our solution for $y^*$ minimizes the loss function.